
\font\bfuno=cmbx12
\font\titolo=cmss17

\font\smallcaps=cmcsc10 

\def\Ham{{\cal H}} 
\def\V{{\cal V}}
\def\C{{\cal C}} 
\def\D{{\cal D}}
\def\A{{\cal A}}
\def\K{{\cal K}}
 
\def\Dif{\,\hbox{\sf D}}
\def\Rfi{{I\mskip-7mu R}}

\def\sss{\scriptscriptstyle}

\magnification1200

{\voffset=6.8truecm
\hoffset=1.5truecm
\hsize=14truecm
\vsize=19truecm
\nopagenumbers
 
{\baselineskip24truept
\centerline{{\titolo Reducing Scattering
Problems}} 

\centerline{{\titolo under  Cone Potentials  to
Normal Form}} 

\centerline{{\titolo by Global Canonical
Transformations}}
}
               
\medskip\smallskip                                                            

\centerline{{\smallcaps Gianluca Gorni}}
\centerline{{\it Universit\`a di Udine}}
\centerline{{\it Dipartimento di Matematica 
             e Informatica}}
\centerline{{\it
             via Zanon 6,
             33100  Udine, Italy}}
\medskip
\centerline{{\smallcaps Gaetano Zampieri}}
\centerline{{\it Universit\`a di Padova}}
\centerline{{\it Dipartimento  di Matematica
        Pura e Applicata}}
\centerline{{\it
              via   Belzoni 7, 
              35131 Padova, Italy}}

\vfil

\centerline{May, 1989}

\bigskip
 
{\bf Abstract.}
We introduce a class of Hamiltonian
scattering systems which can be reduced to the
\it ``normal form'' \rm \  
$\dot P=0$, $\dot Q=P$, by means of a
global \it canonical  transformation \rm $\ 
(P,Q)=\A(p,q),\   p,q\in \Rfi^n\,$,  defined 
through \it asymptotic \rm properties of the
trajectories.

These systems are obtained requiring certain
geometrical conditions on 
$\dot p=-\nabla\V(q)$, $\dot q=p$,
where $\V$ is a bounded below \it ``cone
potential''\rm, i.e., the force $-\nabla\V(q)$
always belongs to a closed convex cone which
contains no straight lines.

We can deal with very different asymptotic
behaviours of the potential and the potential
can undergo small perturbations in any arbitrary 
compact set without losing the existence and 
the properties of $\A$.

\vfil

\centerline{This research was supported by the
     {\it Ministero della Pubblica Istruzione}
            and by the C.N.R.}

\eject
}

scaled\magstep1 \font\bfuno=cmbx12
\font\titolo=cmss17

\font\smallcaps=cmcsc10
\font\sf=cmss10
\font\smallcaps=cmcsc10

\def\Ham{{\cal H}} 
\def\V{{\cal V}}
\def\C{{\cal C}} 
\def\D{{\cal D}}
\def\A{{\cal A}}
\def\K{{\cal K}}
 
\def\Dif{\,\hbox{\sf D}}
\def\Rfi{{I\mskip-7mu R}}

\def\sss{\scriptscriptstyle}

\pageno=2

\centerline{{\bfuno 1. Introduction}}

\bigskip\medskip

This paper presents new results within the
theory developed in~[3] on the Hamiltonian
systems
$$\dot p=-\nabla \V (q)\,,\qquad 
  \dot q=p\,,\qquad 
  p,q\in\Rfi^n\,.
  \eqno(1.1)$$
$\V$ is assumed to be a {\it cone potential},
that is, the force $-\nabla \V (q)$ is always 
in a closed convex cone which contains 
no lines. If we also assume that $\V$ is 
bounded below (so that in particular the 
solutions globally exist), then the 
velocity $p$ has a finite limit 
$p_{\sss\infty}$ as time goes to $+\infty$. 
This remarkably simple fact was showed by 
Gutkin in~[4]. 

We denote by $t\mapsto(p(t,\bar p,\bar q),
q(t,\bar p,\bar q))$ the solution to~(1.1) 
with $(\bar p,\bar q)$ as  initial data:
$p(0,\bar p,\bar q)=\bar p$,
$q(0,\bar p,\bar q)=\bar q$.
The function
$$p_{\sss\infty}(\bar p,\bar q)
  :=\lim_{t\to +\infty} p(t,\bar p,\bar q)
    \in \Rfi^n\,,
    \eqno(1.2)$$
is trivially constant along the motions
solving~(1.1). 

In~[3] we found sufficient conditions on~$\V$ 
for the components of $p_{\sss\infty}$ to be 
$C^k$  ($2\le k\le +\infty$) first integrals, 
independent and pairwise in involution. 
We could thereby construct a class of integrable
Hamiltonian systems with cone potentials.

The present paper extends the investigation 
from the asymptotic velocities $p_{\sss\infty}$
to the limits, sometimes referred to as 
asymptotic phases,
$$a_{\sss\infty}(\bar p,\bar q)
  :=\lim_{t\to +\infty}
  \bigl(q(t,\bar p,\bar q)-
    t\, p(t,\bar p,\bar q)\bigr)
    \in \Rfi^n\,.
    \eqno(1.3)$$
We find sufficient conditions which,
in particular, guarantee the existence and 
smoothness of these limits as functions 
of the initial data $(\bar p,\bar q)$ 
(Proposition~2.2). 

A geometric interpretation of
$a_{\sss\infty}$ follows from the fact that
the limit of $(q(t)-t\, p_{\sss\infty})$
as $t\to +\infty$ is $a_{\sss\infty}$ too
(Proposition~2.2), so that the motion $t\mapsto
q(t,\bar p,\bar q)$, as $t\to+\infty$, is
asymptotically rectilinear uniform:
$$q(t,\bar p,\bar q)= 
  a_{\sss\infty}(\bar p,\bar q)+
  t\,p_{\sss\infty}(\bar p,\bar q)
  +o(1)\qquad
  \hbox{as $t\to+\infty$,}$$
and, in particular, the straight line
$\{a_{\sss\infty}(\bar p,\bar q)+
\xi\,p_{\sss\infty}(\bar p,\bar q)
\;;\; \xi\in\Rfi\}$ is an {\it asymptote} for
the trajectory.

We call $(p_{\sss\infty}(\bar p,\bar q),
a_{\sss\infty}(\bar p,\bar q))$ {\it
the asymptotic  data}, and 
$$\A\colon (\bar p,\bar q)\mapsto
  (p_{\sss\infty}(\bar p,\bar q),
  a_{\sss\infty}(\bar p,\bar q))
  \eqno(1.4)$$ 
{\it the asymptotic map}. Of course we can 
consider the corresponding {\it asymptotic map 
in the past}
$$\A_-\colon (\bar p,\bar q)\mapsto
  (p_{\sss-\infty}(\bar p,\bar q),
  a_{\sss-\infty}(\bar p,\bar q))$$ 
by taking the limits in~(1.2) and~(1.3) 
as time goes to~$-\infty$, and finally the 
{\it scattering map} $\A\circ \A_-^{-1}$.

In this paper we strengthen the hypotheses 
of~[3] and, as a main result (Theorem~2.3), 
we prove that: 

\medskip\sl
 
\item{a)} $\A$ is a global $C^k$-diffeomorphism
  (\/$2\le k\le+\infty$), 
  whose exact range has a simple geometric 
  description;
 
\smallskip

\item{b)} $\A$ is a  canonical
  transformation;
    
\smallskip

\item{c)} $\A$ transforms the Hamiltonian 
  $\Ham:={1\over2}|p|^2+\V(q)\,$, which 
  defines~(1.1), into 
  $$\K(P,Q)\colon=\Ham \circ
    \A^{-1}(P,Q)={1\over2}|P|^2\,, 
    \eqno(1.5)$$
  (up to a trivial additive constant),
  that is, $\A$ transforms~(1.1) into the 
  simple linear form 
  $$\dot P=0\,,\qquad \dot Q=P\,,
    \eqno(1.6)$$ 
  which is precisely what we mean by ``normal''
  in the title

\medskip\rm

\noindent
(see Appendix I for the definition of
 canonical transformations and 
the properties that we use; remark that this
definition  is more restrictive than the
one adopted by several authors). 

Of course we have symmetrical results for
$\A_-$. Thus, the scattering map 
$\A\circ \A_-^{-1}$ is proved to be a 
 canonical transformation 
too, and we can exhibit its exact domain 
and range (see the Remark following Theorem~2.3). 
 
Let us also point out that the canonical
variables $(P,Q)$, introduced by the asymptotic
map $\A$, are similar to the celebrated 
{\it action-angle variables} $I$, $\phi$ 
($\phi$ defined mod~$2\pi$), which transform 
Hamilton equations into  
$$\dot I=0\,,\qquad \dot \phi=\omega(I)\,.
  \eqno(1.7)$$ 
---see [2], Chapter~4.
Of course, equations~(1.6) describe a scattering
system ($Q$ is not defined mod~$2\pi$), 
while~(1.7) represents oscillations.
 
We think that a noteworthy (and probably new)
property is the following: the complete
integrability of our systems, as well as the
the existence of the  variables
$(P,Q)=\A(p,q)$ yielding~(1.6), are
{\it persistent under any small perturbation 
of the potential in an arbitrary compact set} 
(Persistence Theorem~2.4).

\bigskip
\line{\hfil * \qquad * \qquad * \hfil }
\bigskip

Now, after the short rewiew of the results, 
let us briefly outline the nature of our 
hypotheses and highlight some of the crucial 
points of the proofs. 

We always assume on the bounded below 
cone potential $\V$ some global properties, 
to be discussed later, and a certain  
``decay law'' for the gradient $\nabla\V$
(Hypo\-theses~2.1, iv), whence the existence 
and continuity of the asymptotic map~$\A$. 

To achieve the differentiability of~$\A$, 
we first consider potentials with 
{\it exponential} asymptotic behaviour of the
partial derivatives (Hypotheses~2.1, v). This
assumption, together  with Gronwall a priori
estimates on the solutions  of the linear
variational equations, permits to prove that the
limits~(1.2) and~(1.3) are in the $C^k$ norm,
locally in $(\bar p, \bar q)$. 

As in~[3], we can deal with {\it
nonexponential} asymptotic decays of the 
derivatives of~$\V$ by the use of the side 
hypotheses of {\it convexity} on $\V$ 
and a kind of {\it monotonicity}  in the  
Hessian matrix of $\V$ (Hypotheses 2.1, vi).  
These assumptions permit the use of some  {\it
Liapunov  functions} which yield  a priori
estimates much sharper than the  mere Gronwall
ones. 

The {\it global} conditions that we mentioned
(Hypotheses 2.1 ii,iii) are used in~[3] first
to guarantee that the asymptotic velocity 
always belongs to the {\it interior} of the 
convex closed cone $\D$ defined as the dual of 
the cone $\C$ spanned by the forces, i.e., 
$$\eqalignno{
  \C:={}&\biggl\{ -\sum_{\alpha\in I}
  \lambda_{\alpha}\,\nabla \V (q_{\alpha})\;:
  \;\emptyset\ne I\hbox{ finite set, }
  \lambda_{\alpha}\ge0,\  q_{\alpha}\in \Rfi^n
  \;\forall\alpha\in I\biggr\}\,,
  &(1.8)\cr
  \D:={}&\bigl\{v\in \Rfi^n\, :\,
  w\, \cdot \, v\, \ge\, 0\,
  \quad \forall w\in \C\bigr\}\,,
  &(1.9)\cr}$$
and then to prove the following crucial {\it 
locally uniform} estimate on the trajectories:
for every $(\bar p_{\sss 0},\bar q_{\sss 0})$ 
there exist $\gamma>0$, $t_{\sss 0}\in\Rfi$ 
and a neighbourhood $U$ of $(\bar p_{\sss 0},
\bar q_{\sss0})$ such that
$$p(t,\bar p,\bar q)\in \D^{\circ}\,, \qquad
  \hbox{dist}\bigl(p(t,\bar p,\bar q),\partial\D
  \bigr)\ge\gamma\,, 
  \eqno(1.10)$$
for every $t\ge t_{\sss 0}$ and every 
$(\bar p,\bar q)\in U$. 

These global conditions on the potential, in
particular, require that the scalar product 
between any two forces is nonnegative. So, 
in our framework, {\it the cone $\C$ has 
width not larger than~$\pi/2$}. 

In Section~3 we give some examples where our
theory applies. Corollary 3.3 is about
potentials with inverse $r$-power for arbitrary
$r>1$. The complete integrability holds for
arbitrary $r>0$, as it is shown in~[3]
(in~[3], Section~1, we also refer to some
papers of Calogero, Marchioro and Moser who
discovered an analytically integrable system
where the potential has $r=2$ and cone of the
forces wider than $\pi /2$).

Our assumptions hold for a class of Toda-like 
systems (Corollary 3.4), which are defined 
through finite sums of exponentials. The
complete integrability of these systems was
proved in~[9]. Both Corollaries~3.3 and~3.4 are
consequences of Proposition~3.2, where more
general functions are considered.
  
These examples of cone potentials, as well as 
all the ones considered in the literature 
(as far as we know), have polyhedrical cone 
$\C$ of the forces.  Our Section~3 starts by 
studying a simple example where $\C$ is 
{\it not polyhedrical} (namely, it is circular;
Proposition~3.1). In~fact, the present approach 
does not exploit such additional  structures
of~$\V$ as being finite sum of one-dimensional
functions (in the sense of Proposition~3.2). 
Therefore it may be called theory of ``cone'' 
potential with width~$\le\pi/2$.

\bigskip\bigskip
\line{\hfil * \qquad * \qquad * \hfil }
\bigskip\null\goodbreak
\bigskip

The complete solution of the scattering problem 
associated with the non-periodic Toda lattice, 
and many related topics, was given by Moser 
in~[8] by methods different from ours. 
These systems are physically very interesting 
since they describe the dynamics of finitely 
many particles on the line under the influence
of  pairwise interactions with exponential
potential. In such case the cone spanned by the
forces is wider than~$\pi/2$ and there is
analytic integrability. 

Gutkin in~[5] and~[6] considered generically
the problem of the existence of the limits 
in~(1.3), and he studied the global existence 
(i.e. for every $(\bar p, \bar q)$) 
in~[7], specially in connection with the 
problem of the scattering of particles
in the line with pairwise interactions. 
In the title of~[7], as well as in the whole 
paper,  a ``regular trajectory'' is a trajectory 
for which $p_{\sss\infty}\in \D^{\circ}$ and the 
limit in~(1.3) exists. In Gutkin's framework 
the cone spanned by the forces has arbitrary 
width~$<\pi$, but no estimates similar to~(1.10)
are given. Whether they hold anyway, and whether 
the methods of the present paper can be adapted 
to the systems considered in~[7] are open
problems. We guess they are worth studying even
in very  particular contexts like low dimensions
and  special forms of the potential. 

   \vfill\eject

\centerline{{\bfuno 2.  Existence and Regularity
of  the Asymptotic Map}}

\bigskip

Given a smooth function $\V\colon\Rfi^n\to\Rfi$,
we will denote by~$\nabla\V$ its gradient, as a
column vector, and $\Dif^m\V$~will be its $m$-th
differential, regarded as a multilinear map from
$(\Rfi^n)^{m-1}$ into~$\Rfi^n$, endowed with the
norm
$$\|\Dif^m\V(q)\|:=\sup\{
  |\Dif^m\V(q)(x^{(1)},\ldots,x^{(m-1)})|\;:\;
  x^{(i)}\in\Rfi^n,\;|x^{(i)}|\le1\}.$$
Throughout this section the function $\V$ is
assumed to be defined on all of~$\Rfi^n$, but
only for convenience of notation. Everything
runs just as well if the domain is a set of the
form~$q+\D^\circ$. Remind that, given a cone
$\C\subset\Rfi^n$, we define the dual cone~$\D$
as
$$ \D:=\bigl\{v\in \Rfi^n\, :\,
   w\cdot v \ge 0\,
   \quad \forall
   w\in \C\bigr\}\,.$$

\bigskip

{\bf Hypotheses 2.1} \sl $\V$ is a function in
$C^{m+1}(\Rfi^n)$, $m\ge2$. Let $\C$~be the the
convex cone generated by $-\nabla\V$, and
$\D$~be the dual cone of~$\C$. We assume i)
to~iv): 

\medskip

\item{i)} $\inf\V>-\infty$;

\smallskip

\item{ii)} for each $E>0$ 
there exists a $q_{\sss E}\in\Rfi^n$ such that
$$q\in \Rfi^n\backslash
  (q_{\sss E}+\D)
  \quad\Rightarrow\quad 
  \V(q)\ge E\, \hbox{;}$$

\item{iii)} for each 
$q^\prime\,,\,q^{\prime\prime}\in\Rfi^n$ 
such that 
$q^{\prime\prime}\in  q^\prime+\D$,
and for each
$v\in\bar\C\backslash\{0\}$ 
there exists 
$\varepsilon>0$ such that
$$\Bigl(\; q\in q^\prime+\D
  \quad\hbox{and}\quad 
  q\cdot v\le
  q^{\prime\prime}\cdot v \;\Bigr)
  \quad\Rightarrow\quad
  -\nabla\V(q)\cdot v\ge\varepsilon\,\hbox{;}$$

\item{iv)} there exist a vector
$q_{\sss0}\in\Rfi^n$ and a
weakly decreasing function
$h_{\sss0}\colon\Rfi_+\to\Rfi$  such that 
$\int_0^{+\infty}xh_{\sss0}(x)\,dx<+\infty$ and
$$q\in q_{\sss0}+\D
  \quad\Rightarrow\quad
  |\nabla\V(q)|\le
  h_{\sss0}\Bigl(\,
    \hbox{{\rm dist}}\bigl(q,q_{\sss0}
                    +\partial\D\bigr)\,
  \Bigr)\hbox{;}$$

\medskip

\noindent
and either one of v) and vi):

\medskip

\item{v)} there exist
$q_{\sss1},\ldots,q_{\sss m}\in\Rfi^n$,
$A_{\sss1},\ldots, A_{\sss m}\ge0$,
$\lambda_{\sss1},\ldots,\lambda_{\sss m}>0$ such
that  
$$q\in q_{\sss i}+\D\quad\Rightarrow\quad
  \|\Dif^{i+1}\V(q)\;\|\le
  A_{\sss i}\exp\Bigl(-\lambda_{\sss i}
  \hbox{{\rm dist}}\bigl(
  q,q_{\sss i}+\partial\D\bigr)
  \,\Bigr)\,\hbox{,}$$

\item{vi)} there exist 
$q_{\sss1},\ldots,q_{\sss m}\in\Rfi^n$ 
and weakly decreasing functions
$h_{\sss1},\ldots h_{\sss m}\colon\Rfi_+\to\Rfi$
such that 
\itemitem{$\bullet$} $\V$ is convex on
  $q_{\sss1}+\D$; 
\itemitem{$\bullet$} for all
  $q^\prime,\;q^{\prime
  \prime}\in q_{\sss1}+\D$ and all $z\in\Rfi^n$
  we have
  $$q^{\prime\prime}\in q^\prime+\D\quad
    \Rightarrow\quad
    \Dif^2\V(q^{\prime\prime})z\cdot z\le
    \Dif^2\V(q^\prime)
    z\cdot z\hbox{;}$$
\itemitem{$\bullet$} for all $i$,
  $\int_0^{+\infty}\mskip-7mu 
  x^{i+1}h_{\sss i}(x)dx<+\infty$ and
  $$q\in q_{\sss i}+\D\quad\Rightarrow\quad
    \|\Dif^{i+1}\V(q)\|\le h_{\sss i}\Bigl(\,
    \hbox{{\rm dist}}\bigl(q,q_{\sss i}+
    \partial\D\bigr)\,\Bigr)\hbox{;}$$

\rm

\goodbreak
\bigskip

For a potential $\V$ satisfying the previous
assumptions, we consider the Hamiltonian system
$$\dot p=-\nabla\V(q)\,,\qquad 
  \dot q=p\,,
  \eqno(2.1)$$
the associated flow
$$\Phi^t(\bar p,\bar q)=
  {p(t,\bar p,\bar q)\choose
  q(t,\bar p,\bar q)}\,.
  \eqno(2.2)$$
and the mappings
$$\A_t(\bar p,\bar q):=
  {p(t,\bar p,\bar q)\choose
   q(t,\bar p,\bar q)-t\,p(t,\bar p,\bar q)}=
  \left(\matrix{I_n&0\cr-tI_n&I_n\cr}\right)
  \Phi^t(\bar p,\bar q)\,.
  \eqno(2.3)$$
We are interested in the asymptotic map
$${p_{\sss\infty}(\bar p,\bar q)\choose
  a_{\sss\infty}(\bar p,\bar q)}=
  \A(\bar p,\bar q):=\lim_{t\to+\infty}
  \A_t(\bar p,\bar q)\,.
  \eqno(2.4)$$

\goodbreak
\bigskip

{\bf Proposition 2.2} \sl
Suppose that the Hypotheses 2.1 hold.
Then the the asymptotic map $\A$ exists, and the
convergence is locally in the
$C^m$ norm. The functions
$p_{\sss\infty}$ and $a_{\sss\infty}$ verify
$$\eqalign{
  p_{\sss\infty}(\Phi^t(\bar p,\bar q))&{}=
  p_{\sss\infty}(\bar p,\bar q)\,\hbox{,}\cr
  a_{\sss\infty}(\Phi^t(\bar p,\bar q))&{}=
  a_{\sss\infty}(\bar p,\bar q)+t\,
  p_{\sss\infty}(\bar p,\bar q)\,\hbox{,}\cr}
  \eqno(2.5)$$
that is to say,
$$\A\circ\Phi^t=
  \left(\matrix{I_n&0\cr tI_n&I_n\cr}\right)
  \A\,.
  \eqno(2.6)$$
Moreover,
$$\eqalign{a_{\sss\infty}(\bar p,\bar q)={}&
  \lim_{t\to+\infty}\bigl(
  q(t,\bar p,\bar q)-
     t\,p(t,\bar p,\bar q)\bigr)=\cr
  ={}&\lim_{t\to+\infty}\bigl(
  q(t,\bar p,\bar q)-
     t\,p_{\sss\infty}(\bar p,\bar q)\bigr).\cr}
  \eqno(2.7)$$
\rm

\goodbreak
\bigskip

{\bf Proof. } Most of the arguments needed for
the proof are an adaptation of
the ones given in~[3]. The starting point is
the identity
$$\A_t(\bar p,\bar q)=
  {\bar p\choose\bar q}+
  \int_0^t{-I_n\choose sI_n}
  \nabla\V(q(s,\bar p,\bar q))\,ds\,,
  \eqno(2.8)$$
which follows from the Hamilton equations. The
reasoning of~[3], Sections~4 and~5, proves that
we can go to the limit in the formula, and the
convergence is locally uniform. If we let
$\tau\to+\infty$ in the next identity
$$\left(\matrix{I_n&0\cr-\tau I_n&I_n\cr}\right)
  \Phi^\tau\circ\Phi^t=
  \left(\matrix{I_n&0\cr-\tau I_n&I_n\cr}\right)
  \Phi^{\tau+t}=
  \left(\matrix{I_n&0\cr tI_n&I_n\cr}\right)
  \left(\matrix{I_n&0\cr-(\tau+t)I_n&I_n\cr}
  \right)
  \Phi^{\tau+t}$$
we get formulas~(2.6) and~(2.5).\hfil\break  
The proofs of~[3], Sections~6, 7 and~8, can be
adjusted to obtain that the convergence in
formula~(2.4) is locally~$C^m$.\hfil\break To end
with, remark the two following identities:
$$\eqalign{q(t,\bar p,\bar q)-
     t\,p(t,\bar p,\bar q)={}&
  \bar q+\int\mskip-10mu\int_{0\le r\le s\le t}
  \nabla\V(q(s,\bar p,\bar q))\,dr\,ds\,,\cr
  q(t,\bar p,\bar q)-
     t\,p_{\sss\infty}(\bar p,\bar q)={}&
  \bar q+\int\mskip-10mu\int_{{0\le r\le t\atop
                   r\le s}}
  \nabla\V(q(s,\bar p,\bar q))\,dr\,ds\,.\cr}$$
The proof of the existence of $\A$ actually
amounted to showing that the function $(r,s)
\mapsto|\nabla\V(q(s,\bar p,\bar q))|$ is
integrable on the set $\{(r,s)\in\Rfi^2\;:\;
0\le r\le s\}$, whence the equality~(2.7).

\line{\hfil$\diamondsuit$}
\goodbreak
\bigskip

{\bf Theorem 2.3} \sl Suppose that
Hypotheses~2.1 hold, namely, i) to iv) and
either v) or~vi). Then the asymptotic map~$\A$
exists, it is a $C^m$ diffeomorphism of 
$\Rfi^n\times\Rfi^n$ onto
$\D^\circ\times\Rfi^n$ and it is a 
canonical transformation. The transformed
Hamiltonian is
$$\K(P,Q)\colon=\Ham \circ
  \A^{-1}(P,Q)={1\over2}|P|^2+\inf\V\,,
  \eqno(2.9)$$
and the associated system is
$$\dot P=0\,,\qquad \dot Q=P\,.
  \eqno(2.10)$$
\rm

{\bf Proof.} We refer again to the results and
methods developed in~[3]. Proposition~4.3
of that paper shows in particular that the range
of~$\A$ is contained in $\D^\circ
\times\Rfi^n$.\hfil\break 
As in~[3], Propositions~5.3, 6.4 and~7.5,
given~$\gamma$, there exists $\tilde
q_\gamma\in\Rfi^n$ such that, if we define the
set $A_\gamma$ as 
$$A_\gamma:=\{(\bar p,\bar q)\;:\;
  \bar p\in\D^\circ,\;
  \hbox{dist}(\bar p,\partial\D)>\gamma\,,\;
  \bar q\in\tilde q_\gamma+\D\}\,,$$
then
$$(\bar p,\bar q)\in A_\gamma
  \quad\Rightarrow\quad
  \Bigl|\A(\bar p,\bar q)-{\bar p\choose\bar q}
  \Bigr|+
  \Bigl\| \Dif\A(\bar p,\bar q)-
  I_{2n}\Bigr\|\le{1\over2}\,.
  \eqno(2.11)$$
Let $(\bar p_{\sss0},\bar q_{\sss0})\in\D^\circ
\times\Rfi^n$,  $t_{\sss0}\in\Rfi$ be such that
$(\bar p_{\sss0},\bar q_{\sss0}+t_{\sss0}\bar
p_{\sss0})\in A_\gamma$, dist$(\bar
q_{\sss0}+t_{\sss0}\bar p_{\sss0},
\tilde q_\gamma+\partial\D)\ge1/2$,
where $2\gamma:={}$dist$(\bar p_{\sss0},
\partial\D)$. Then the mapping 
$${\bar p\choose\bar q}\quad\mapsto\quad
  {\bar p_{\sss0}\choose
  \bar q_{\sss0}+t_{\sss0}\bar p_{\sss0}}+
  {\bar p\choose\bar q}-
  \A(\bar p,\bar q)$$
is a contraction of the closed ball 
$$\{(\bar p,\bar q)\;:\;
  \bigl|
  (\bar p,\bar q)-(\bar p_{\sss0},
  \bar q_{\sss0}+t_{\sss0}\bar p_{\sss0})
  \bigr|
  \le1/2\}$$
into itself. The corresponding fixed point 
$(\bar p_{\sss0}^\prime,\bar q_{\sss0}^\prime)$
verifies
$$\A(\bar p_{\sss0}^\prime,
     \bar q_{\sss0}^\prime)=
  {\bar p_{\sss0}
  \choose
  \bar q_{\sss0}+t_{\sss0}
  \bar p_{\sss0}}$$
and therefore
$$\A\bigl(\Phi^{-t_0}(\bar p_{\sss0}^\prime,
     \bar q_{\sss0}^\prime)\bigr)=
  \left(\matrix{I_n&0\cr-t_{\sss0}I_n&I_n\cr}
  \right)
  {\bar p_{\sss0}
  \choose
  \bar q_{\sss0}+t_{\sss0}
  \bar p_{\sss0}}=
  {\bar p_{\sss0}
  \choose
  \bar q_{\sss0}}.$$
This proves that $\A$ is onto $\D^\circ\times
\Rfi^n$.\hfil\break
To prove that $\A$ is a local diffeomorphism, 
we start remarking that $\A_t$ is canonical. In
fact, it is the composition of the
transformations
$$\left(\matrix{I_n&0\cr-tI_n&I_n\cr}
  \right)\quad\hbox{ and }\quad
  \Phi^t$$
which are canonical (the latter through a
general theorem and the former with a direct
computa\-tion---see the Appendix~I). From the
identity (see (4.2) in Appendix I)
$$(\Dif\A_t)\left(\matrix{0&-I_n\cr I_n&0\cr}
  \right)(\Dif\A_t)^T=
  \left(\matrix{0&-I_n\cr I_n&0\cr}
  \right),$$
going to the limit as $t\to+\infty$, we obtain
$$(\Dif\A)\left(\matrix{0&-I_n\cr I_n&0\cr}
  \right)(\Dif\A)^T=
  \left(\matrix{0&-I_n\cr I_n&0\cr}
  \right),$$
which at once gives that $\Dif\A$ is
nonsingular, and will yield also that $\A$ is
canonical, as soon as we prove that it is
one-to-one.\hfil\break
To show this, we can start by
noticing that $\A$~is certainly one-to-one on
$A_\gamma$ for any $\gamma>0$, because
on the {\it convex} set $A_\gamma$ the symmetric
part of the Jacobian matrix $\Dif\A$ is positive
definite (for the proof of this simple and
well-known global injectivity result, see the
Appendix~II). But for any couple of initial data
$x_{\sss0}\ne x_{\sss1}\in\Rfi^n\times\Rfi^n$, 
the points $\Phi^t(x_{\sss0})$ and 
$\Phi^t(x_{\sss1})$ belong to the set $A_\gamma$
from a certain time $t_{\sss1}$ on, where
$2\gamma$~is the smaller between the distance of
$p_{\sss\infty}(x_{\sss0})$ and 
$p_{\sss\infty}(x_{\sss1})$ from the boundary
of~$\D$. Since $\Phi^t$ is always one-to-one, we
surely have $\Phi^{t_1}(x_{\sss0})\ne\Phi^{t_1}
(x_{\sss1})$, and consequently 
$\A(\Phi^{t_1}(x_{\sss0}))\ne\A(\Phi^{t_1}
(x_{\sss1}))$. The conclusion $\A(x_{\sss0})
\ne\A(x_{\sss1})$ comes from the identity~(2.6)
rewritten in the form
$$\A=
  \left(\matrix{I_n&0\cr t_{\sss1}I_n&I_n\cr}
  \right)^{-1}
  \A\circ\Phi^{t_1}.
  \eqno(2.12)$$
Finally,  (2.10) comes from (2.5) by derivation
in the time, reminding that 
$(P,Q):=\A=(a_{\sss\infty},p_{\sss\infty})$.
The Hamiltonian of the transformed system (2.10)
is $\K(P,Q)={1\over2}|P|^2$ up to a trivial
additive constant. Actually this constant is
$\inf\V\,$  by means of Corollary~4.6 in [3].

\line{\hfil$\diamondsuit$}
\goodbreak
\bigskip

{\bf Remark.} \sl Let $\A_-$ be defined as $\A$
but with the limits as $t\to -\infty$ instead of
$t\to+\infty$. Then, by reversing the time, we
obtain for $\A_-$ similar results. So the {\it
scattering map} $\A\circ \A_-^{-1}$ exists and
it is a  canonical transformation 
with domain $-\D^\circ\times \Rfi^n$ and range 
$\D^\circ\times \Rfi^n$. \rm

\goodbreak
\bigskip

The next theorem uses the same arguments as for
Theorem~9.3 of~[3], so that we omit the proof.

\bigskip

{\bf Theorem 2.4 (Persistence) } \sl
Suppose that $\V$ verifies Hypotheses~2.1. Let
$K\subset\Rfi^n$ be compact.  Then there exists
an $\varepsilon>0$ with the following property.
If $f\colon\Rfi^n\to\Rfi$ is a $C^{m+1}$ function
with support in $K$ and  
$$\sup|\nabla f|<\varepsilon\,,
  \eqno(2.13)$$
then for the system whose Hamiltonian $H$ is 
$$H(p,q):={1\over2}|p|^2+\V(q)+f(q)
  \eqno(2.14)$$
all the claims of Proposition~2.2 and
Theorem~2.3 hold. \rm

   \vfill\eject

\centerline{{\bfuno 3. Examples}}
\bigskip

Consider the closed convex cone in $\Rfi^3$
defined by
$$\D:=\{(x,y,z)\in\Rfi^3\;:\;
  z\ge\sqrt{x^2+y^2}\}\,.
  \eqno(3.1)$$
The function
$$f(x,y,z):={z^2-x^2-y^2\over2z}
  \eqno(3.2)$$
is continuous and positive on the interior
of~$\D$, and its level sets are single sheets of
two-sheeted hyperboloids of revolution:
$${z^2-x^2-y^2\over2z}=c\quad\iff\quad
  (z-c)^2-x^2-y^2=c^2,\quad z\ne0.$$
The function $f$ is closely related to the
distance of a point of~$\D^\circ$ from the
boundary of~$\D$: 
$$0<f(x,y,z)\le z-\sqrt{x^2+y^2}=\sqrt2
  \hbox{ dist}\bigl((x,y,z),\partial\D\bigr)\le
  2f(x,y,z)\,.
  \eqno(3.3)$$
In particular, $f$ vanishes on the boundary
of~$\D$. The gradient of~$f$
$$\nabla f(x,y,z)=\Bigl(-{x\over z}\,,\,
  -{y\over z}\,,\,{1\over2}+{x^2+y^2\over2z^2}
  \Bigr)
  \eqno(3.4)$$
spans a cone $\C$ which happens to coincide with
the interior of~$\D$. The $n$-th differential
$\Dif^n f(x,y,z)$ is homogeneous of degree~$1-n$
with respect to $(x,y,z)$.

The one-variable function
$$\varphi(r):={e^{-r}\over r}\,,\qquad
  r>0\,,
  \eqno(3.5)$$
is positive, $\lim_{r\to0+}\varphi(r)=+\infty$,
$\lim_{r\to+\infty}\varphi(r)=0$, and the $n$-th
derivative $\varphi^{(n)}$ can be expressed as
$$\varphi^{(n)}(r)=(-1)^nP_n\Bigl({1\over r}
  \Bigr)e^{-r},
  \eqno(3.6)$$
where $P_n$ is a nonzero polinomial with
nonnegative coefficients. In particular
$$r\ge1\quad\Rightarrow\quad
  |\varphi^{(n)}(r)|\le P_n(1)e^{-r}.
  \eqno(3.7)$$

Let us now define on $\D^\circ$ the following
potential
$$\V(x,y,z):=\varphi(f(x,y,z))\,.
  \eqno(3.8)$$
This is a $C^\infty$ function, and $-\nabla\V$
spans the cone~$\C$. We are going to verify the
Hypo\-theses~2.1 i) to~v).

It is obvious that $\V>0$. Next, for a given
$E>0$, choose $\varepsilon>0$ such that
$0<r\le\varepsilon\allowbreak
\Rightarrow\varphi(r)\ge E$.
Then
$$(x,y,z)\in\D^\circ\backslash\bigl(
  (0,0,\varepsilon)+\D\bigr)
  \quad\Rightarrow\quad
  0<f(x,y,z)\le\varepsilon
  \quad\Rightarrow\quad
  \V(x,y,z)\ge E\,,
  \eqno(3.9)$$
and  ii) is settled. Verifying iii) is more
complicated. Because of the symmetry with
respect to the $z$-axis, instead of a generic
vector $v\in\bar\C\backslash\{0\}$, we can just
take $(\alpha,0,1)$, with $-1\le\alpha\le1$.
What we need is a positive lower bound on
$$-\nabla\V(x,y,z)\cdot(\alpha,0,1)=
  -\varphi^\prime(f(x,y,z))\nabla f(x,y,z)
  \cdot(\alpha,0,1)
  \eqno(3.10)$$
when $(x,y,z)$ belongs to the set
$$M:=\{(x,y,z)\in\Rfi^3\;:\;
  a+\sqrt{x^2+y^2}\le z\le b-\alpha x\}
  \eqno(3.11)$$
for $0<a\le b$. The $y$ variable can be dispensed
with, first noticing that $M$ is contained in
$\{a+|x|\le z\le b-\alpha x\}$. Next we can
compute
$$\eqalign{
  \nabla f(x,y,z)\cdot(\alpha,0,1)={}&
  \nabla f(x,y,z)\cdot\Bigl(
  {1-\alpha\over2}(-1,0,1)+
  {1+\alpha\over2}(1,0,1)\Bigr)\ge\cr
  \ge{}&
  {1-\alpha\over4}\Bigl({x\over z}+1\Bigr)^2+
  {1+\alpha\over4}\Bigl({x\over z}-1\Bigr)^2.\cr}
  \eqno(3.12)$$
We can write
$$(x,y,z)\in M\quad\Rightarrow\quad\cases{
  {x\over z}+1\ge{a\over b}>0& 
  if $\alpha\le0$,\cr
  \noalign{\smallskip}
  {x\over z}-1\le-{a\over b}<0&
  if $\alpha\ge0$,\cr}
  \eqno(3.13)$$
so that
$$(x,y,z)\in M\quad\Rightarrow\quad
  \nabla f(x,y,z)\cdot(\alpha,0,1)\ge
  {a^2\over4b^2}>0\,.
  \eqno(3.14)$$
Finally, noticing that $(x,y,z)\in M\Rightarrow
a/2\le f(x,y,z)\le b$,
$$(x,y,z)\in M\quad\Rightarrow\quad
  -\nabla\V(x,y,z)\cdot(\alpha,0,1)\ge
  {a^2\over4b^2}\inf\{-\varphi^\prime(r)\;:\;
  a/2\le r\le b\}>0\,.
  \eqno(3.15)$$
All we are left to do is giving an exponential
bound on $\|\Dif^n\V(x,y,z)\|$ in terms of the
distance of $(x,y,z)$ from the boundary of~$\D$.
But $\Dif^n\V(x,y,z)$ is a linear combination of
$n$-th order objects extracted from~$f$ (which
are homogeneous of degree~$1-n$ and continuous
on the half space~$z>0$, hence bounded
on~$\D\cap\{z\ge2\}$), with
$\varphi^{(i)}(f(x,y,z))$ as coefficients. We
can then estimate, using~(3.7),
$$\eqalign{
  (x,y,z)&{}\in(0,0,2)+\D\;\Rightarrow\;
  f(x,y,z)\ge1\Rightarrow\cr
  \Rightarrow{}&\;
  \|\Dif^n\V(x,y,z)\|\le
  A_ne^{-f(x,y,z)}\le
  A_n\exp\Bigl(-{\sqrt2\over2}\hbox{dist}
  \bigl((x,y,z),\partial\D\bigr)\Bigr).\cr}
  \eqno(3.16)$$

\goodbreak
\bigskip

Summing up:

\bigskip

{\bf Proposition 3.1 } \sl The conclusions of
Proposition~2.2 are true for the Hamiltonian
system with potential
$$\V(x,y,z):={2z\over z^2-x^2-y^2}\exp
  \Bigl(-{z^2-x^2-y^2\over2z}\Bigr)
  \eqno(3.17)$$
defined on the set $\D^\circ=
\{(x,y,z)\in\Rfi^3\;:\;z>\sqrt{x^2+y^2}\}$.
The corresponding asymptotic map $\A$
is~$C^\infty$ and the system is
$C^\infty$-integrable. \rm

\goodbreak
\bigskip

The following results are adaptations of the
ones given in~[3], Section~10.

\bigskip

{\bf Proposition 3.2} \sl Let the vectors
$v_{\sss1},\ldots,v_{\sss N}\in\Rfi^n$ be such
that
$$v_\alpha\cdot v_\beta\ge0 \qquad
  \forall\alpha,\beta\,.
  \eqno(3.18)$$
Let the $C^{m+1}$ real functions $f_{\sss1},
\ldots,f_{\sss N}$ be defined either on the
interval $]0,+\infty[$ or on all of~$\Rfi$,
where $\sup f_\alpha=+\infty$,
$f_\alpha>0$, $f_\alpha^\prime<0$. Suppose
moreover that the $f_\alpha$ are integrable on 
an interval of the form $[a,+\infty[$, and that,
for $0\le i\le m+1$ and $x\ge a$, 
$$f_\alpha^{(i)}(x)\cases{<0&if $i$ is odd,\cr
                   >0&if $i$ is even.\cr}
  \eqno(3.19)$$
Assume finally that the $|f_\alpha^{(m+1)}|$
are monotone on~$[a,+\infty[$.
Then the conclusions of Proposition~2.2 hold true
for the Hamiltonian system with the potential 
$$\V(q):=\sum_{\alpha=1}^N
  f_\alpha(q\cdot
  v_\alpha)\,,
  \eqno(3.20)$$
defined on either the set 
$\D^\circ=\{q\in\Rfi^n\;:\;
q\cdot v_\alpha>0\;\forall\alpha\}$
or on all of $\Rfi^n$. \rm

\goodbreak
\bigskip

{\bf Corollary 3.3} \sl Let $v_1,\ldots,v_N\in
\Rfi^n\backslash\{0\}$ be such that $v_\alpha
\cdot v_\beta\ge0$ for all $\alpha,\beta$.
Let $r>1$ and define the potential
$$\V(q):=\sum_{\alpha=1}^N
  {1\over(q\cdot v_\alpha)^r}$$
on the set $\D^\circ=\{q\in\Rfi^n\;\colon\;
q\cdot v_\alpha>0\;\forall\alpha\}$.
Then the conclusions of Proposition~2.2 hold for
the associated Hamiltonian system, and the
asymptotic map~$\A$ is~$C^\infty$. \rm

\goodbreak
\bigskip

{\bf Corollary 3.4} \sl Let $v_1,\ldots,v_N\in
\Rfi^n\backslash\{0\}$ be such that $v_\alpha
\cdot v_\beta\ge0$ for all $\alpha,\beta$, and
let $c_\alpha>0$. Define the potential
$$\V(q):=\sum_{\alpha=1}^N
  c_\alpha\, e^{-q\cdot v_\alpha}$$
on $\Rfi^n$. 
Then the conclusions of Proposition~2.2 hold for
the associated Hamiltonian system, and the
asymptotic map~$\A$ is~$C^\infty$. \rm

   \vfill\eject

\centerline{{\bfuno 4.  Appendix I}}
\bigskip

We consider time-independent $C^2$ Hamiltonian
functions defined in an open domain $\Omega$
of $\Rfi^n\times\Rfi^n$. Such a (real) function
$(p,q)\mapsto H(p,q)$
defines  the Hamiltonian system
$$\dot p=-{\partial H\over\partial q}\,,\qquad
  \dot q={\partial H\over\partial p}\,.
  \eqno(4.1)$$

In this paper {\it  canonical
transformations} are central. We adopt the
following  definition (remark that in the
literature also more general concepts are denoted
with the same name,
and the following transformations are
sometimes called ``completely canonical'').  The
$C^2$-diffeomorphism $Y:\Omega \to Y(\Omega)$ is
said a  canonical transformation if, for \it any
\rm Hamiltonian $H$ as above, the system
 (4.1) is transformed into the Hamiltonian system 
associated to the tranformed Hamiltonian 
$K=H\circ Y^{-1}$.
We see at once that this  is equivalent to
the validity of the following equality at any
point of $\Omega$: 
$$(\Dif Y)
  \left(\matrix{0  &-I_n\cr 
                I_n& 0  \cr}\right)
  (\Dif Y)^T= 
  \left(\matrix{0  &-I_n\cr
                I_n& 0  \cr}\right)\,,
  \eqno(4.2)$$ 
where
$(\Dif Y)$ is the Jacobian matrix, $0$
is the zero $n\times n$ matrix, $I_n$ is the unit
$n\times n$ matrix, and the exponent $T$ means
transposition.
 
The  canonical transformations
constitute a {\it group}.

Let us introduce the notation $(P,Q)\in 
\Rfi^n\times\Rfi^n$ for the transformation $Y$.
An easy theorem says that the preceding
condition is equivalent to requiring that the
differential form 
$$P\,dQ-p\,dq$$
be closed in $\Omega$, or, equivalently,
locally exact at every point $(p,q)\in\Omega$:
$$\sum_{i=1}^n P_i\,{\partial Q_i\over \partial
  p_j}={\partial S\over\partial p_j}\,, \qquad
  \sum_{i=1}^n P_i\,{\partial Q_i\over\partial
  q_j}\,-\, p_j=
  {\partial S\over \partial q_j}\,,\qquad 
  \forall j=1,\ldots,n\,,
  \eqno(4.3)$$ 
for some function $S$ defined near $(p,q)$.

In the present paper we use the following fact:
\it a locally $C^2$ limit of  canonical
transformations is  canonical\rm, 
provided it is a diffeomorphism. This follows
from~(4.2).
  
Let us conclude with the important
theorem stating that, for any fixed value of
the time, {\it the phase flow gives a 
canonical transformation}. 

Let $H\in C^k$, with $k\ge3$. Then, for any
$\lambda \in \Rfi$, the map
$$(\bar p,\bar q)\mapsto 
  \bigl(P^{\lambda}(\bar p,\bar q),
        Q^{\lambda}(\bar p,\bar q)\bigr) :=
  (p(\lambda,\bar p,\bar q),
  q(\lambda, \bar p,\bar q))
  \eqno(4.4)$$
is a $C^{k-1}$ canonical
transformation.\hfil\break
The proof is just verifying that the following
function works for the condition in (4.3):
$$S(\bar p,\bar q)=\int_0^\lambda 
  \biggl(\sum_i p_i(\xi,\bar p,\bar q)\,
  \dot q_i(\xi,\bar p, \bar q)-
  H\bigl(p(\xi,\bar p,\bar q), 
  q(\xi,\bar p,\bar q)\bigr)\,
  \biggr)\,d\xi\,.$$
  
     \vfill\eject

\centerline{{\bfuno 5.  Appendix II}}
\bigskip

In this paper we use the following known 

\bigskip

{\bf Proposition. } \sl
Let $f\colon \Omega \to \Rfi^n$ be a $C^1(\Omega)$
map with $\Omega \subseteq \Rfi^n$ open and
convex. If the quadratic form \ $\xi\mapsto \xi^T
\Dif f(x)\, \xi$\ \  (defined by the Jacobian
matrix) is positive definite at any $x\in\Omega$,
then $f$ is injective.  \rm

\goodbreak
\bigskip

{\bf Proof. }
Let $x\ne y$ be two points of $\Omega$,
and $s$ the segment which joins them
$$s\colon [0,1]\to \Omega,\quad 
  \theta\mapsto\theta x+(1-\theta)y\,.$$
Moreover, consider the map
$$g\colon[0,1]\to\Omega\,,\quad 
  \theta\mapsto
  (x-y)\cdot\bigl(f(s(\theta))-f(y)\bigr)\,.$$ 
The derivative at any $\theta$ is strictly
positive:
$$g'(\theta)=(x-y)^T\Dif f(s(\theta))
  (x-y)>0\,,$$
by the hypothesis of the Proposition. Therefore
the map $g$ is strictly increasing and, in
particular, $g(1)\ne g(0)=0$. Finally 
$$g(1)=(x-y)\cdot \bigl(f(x)-f(y)\bigr)\ne0$$
implies $f(x)\not=f(y)$.

\line{\hfil$\diamondsuit$}

  \vfill\eject

 
\centerline{\bfuno {References}}
\bigskip \frenchspacing

\item{[1]} Arnold, V. I. (1978). 
   {\bf Mathematical methods in classical
   mechanics.} 
   Springer Verlag, Berlin.

\medskip

\item{[2]} Arnold, V. I. (ed.) (1988). 
   {\bf Encyclopaedia of mathematical sciences 3,
     Dynamical Systems III.} 
   Springer Verlag, Berlin.
\medskip

\item{[3]}  Gorni, G., \& Zampieri, G. (1989). 
  {\it Complete integrability for Hamiltonian
  systems with a cone potential}. To appear in 
  {\bf J. Diff. Equat.}

\medskip

\item{[4]} Gutkin, E. (1985). 
   {\it Integrable  Hamiltonians with
   exponential potentials}. 
   {\bf Physica~D~16},
   pp.~398--404,
   North Holland, Amsterdam.

\medskip

\item{[5]}  Gutkin, E. (1985). 
   {\it Asymptotics of  trajectories 
   for cone potential}. 
   {\bf Physica~D~17},
   pp.~235--242.

\medskip 

\item{[6]}  Gutkin, E. (1987). 
   {\it  Continuity of scattering data 
   for particles on the line 
   with directed repulsive interactions}.
   {\bf J.~Math. Phys. 28}, 
   pp.~351--359.

\medskip

\item{[7]} Gutkin, E. (1988).
{\it Regularity of scattering trajectories in
Classical Mechanics}.
{\bf Commun. Math. Phys. 119},
pp.~1--12.

\medskip

\item{[8]} Moser, J. (1975). 
   {\it Finitely many mass points on the line
under the influence of an exponential
potential---An integrable system}. 
   {\bf Proc. Battelle Rencontres, Lec. Notes
in Phys. 38}, 
   pp.~467--497.

\medskip

\item{[9]} Oliva, W.M., 
    \& Castilla, M.S.A.C. (1988).
    {\it On a class of $C^\infty$-integrable 
    Hamiltonian systems}. 
    To appear in the
    {\bf Proc. Royal Soc.
    Edinburgh}
    (in honour of Jack Hale's 60th birthday).

\bigskip 
\centerline{\hbox to3cm{\hrulefill}}

   \bye

\end